\documentclass[11pt]{article}
\usepackage{amsmath,amssymb,amsfonts,fancyhdr,fancybox,
graphics,graphicx,epsfig,calc,color}
\usepackage{multirow}
\usepackage{rotating}
\usepackage{geometry}
\usepackage{empheq}
\graphicspath{{figure/}}

\renewcommand{\baselinestretch}{1.4}
\def\blist#1#2#3
{    \begin{list}{}{
     \setlength{\parsep}{0pt}
     \setlength{\leftmargin}{#1 pt}
     \setlength{\listparindent}{0pt}
     \setlength{\itemindent}{\listparindent}
     \setlength{\labelsep}{#3 pt}
     \setlength{\labelwidth}{\leftmargin}
     \addtolength{\labelwidth}{-\labelsep}
     \addtolength{\labelwidth}{\itemindent}
     \setlength{\rightmargin}{#2 pt}   }   }

\def\elist{\end{list}}

\newtheorem{theorem}{Theorem}
\newtheorem{lemma}{Lemma}

\newtheorem{definition}{Definition}

\newtheorem{example}{Example}

     \def\lm{\lambda}

  \def\bone{{\mathbf 1}} \def\btwo{{\mathbf 2}}

   \def\bI{{\mathbf I}}
  \def\bP{{\mathbf P}} 
   
  \def\bX{{\mathbf X}} 
\def\bZ{{\mathbf Z}}

\def\ba{{\mathbf a}} \def\bb{{\mathbf b}}

 \def\bs{{\mathbf s}} 
  \def\bw{{\mathbf w}}
\def\bx{{\mathbf x}} \def\by{{\mathbf y}} \def\bz{{\mathbf z}}

\def \bbeta{{\boldsymbol{\beta}}}
\def \balpha{{\boldsymbol{\alpha}}}

\def\bmu{{\mathbf \mu}}  \def\hbmu{\widehat{\boldsymbol \mu}}
\def\hmu{\widehat \mu}

\def\bnu{{\mathbf \nu}}  \def\hbnu{\widehat{\boldsymbol \nu}}
\def\hnu{\widehat \nu} \def\tnu{\widetilde \nu}

 \def\bgamma{{\boldsymbol{\gamma}}}
\def\bLambda{{\boldsymbol{\Lambda}}}   \def\bmu{{\boldsymbol{\mu}}}

\def\bnu{{\boldsymbol{\nu}}}   
\def\bphi{{\boldsymbol{\phi}}} 

  \def\bveps{{\boldsymbol{\varepsilon}}}

\def\bPhi{{\boldsymbol{\Phi}}}

 \def\veps{\varepsilon} \def\balpha{{\boldsymbol{\alpha}}}

\def\hbbeta{\widehat{\boldsymbol \beta}}

\def\cW{{\cal W}}

 \def\cB{{\cal B}}  \def\cJ{{\cal J}}

\def\cE{{\cal E}}      
  \def\cR{{\cal R}}  \def\cS{{\cal S}}

\def\II{I\!I}   \def\III{I\!I\!I}
\def\sgn{\mbox{sgn}}  
\def\argmin{\mathop{\rm arg\, min}}

\def\diag{\hbox{diag}}
\def\Var{\hbox{Var}}    
 \def\hbeta{\widehat{\beta}}

\def\real{\mathop{{\rm I}\kern-.2em\hbox{\rm R}}\nolimits}

\def\1overn{\frac{1}{n}}

\def\bel{\begin{eqnarray}\label}  \def\eel{\end{eqnarray}}
\def\bes{\begin{eqnarray*}}  \def\ees{\end{eqnarray*}}

\renewcommand{\baselinestretch}{1.38}

\begin{document}

\topmargin -0.3in \oddsidemargin 0.2in \evensidemargin 0.2in
\baselineskip 9mm
\renewcommand \baselinestretch {1.2}
\title{\bf Joint Linear Trend Recovery Using $\ell_1$ Regularization}
\date{}

\author{\begin{tabular}{c} Xiaoli Gao\footnote{Correspondence: 106 Petty Building, Greensboro, NC 27409. Email: x\_gao2@uncg.edu}\\
\emph{Department of Mathematics and Statistics}\\
\emph{University of North Carolina at Greensboro}\\ \\
Syed Ejaz Ahmed\\
\emph{Department of Mathematics}\\
\emph{Brock University}\\
\end{tabular}}
\titlepage

\maketitle





%
%
%
%
%
%
%
%
%
%
%
\begin{center}
\begin{minipage}{130mm}
\begin{center}{\bf Abstract}
\end{center}
This paper studies
the recovery of a joint piece-wise linear trend from a time series
using $\ell_1$ regularization approach,  called $\ell_1$ trend filtering
(Kim, Koh and Boyd, 2009).
We provide some sufficient conditions under which a
 $\ell_1$ trend filter can be well-behaved in terms of
 mean estimation and change point detection.
The result is two-fold: for the mean estimation, an almost
 optimal consistent rate is obtained;
 for the change point detection,
  the slope change in direction can
  be recovered in a high probability.
  In addition, we show that the
  weak irrepresentable condition, a necessary
  condition for LASSO model to be sign consistent (Zhao and Yu, 2006),
  is not necessary for the consistent change point detection.
The performance of the
$\ell_1$ trend filter is evaluated
by some finite sample simulations studies.
   \end{minipage}
\end{center}

\bigskip
{Keywords:} Change Point, Estimation consistency, $\ell_1$ regularization, Linear trend filtering, Sign consistency.

\newpage
\setcounter{equation}{0}

\newpage
\setcounter{equation}{0}

\section{Introduction}

%

For a naturally-occurring time series with observation $y_t$ and underlying
mean $\mu_t^0$ at $1\le t\le n$, an important issue
is to recover the trend of the underlying mean vector, $\bmu^0=(\mu^0_1,\cdots, \mu^0_n)'$.
Here  $\bmu^0$ often
exhibits various kinds of trends in many real applications.
For example,
in the analysis of DNA sequences,
 $\bmu^0$ is assumed to be piece-wise constant
 (Braun and Muller, 1998; Huang {\it et. al}, 2005). However,
in financial time series,
 $\bmu^0$ is often assumed to be piece-wise linear (Taylor, 2008).
Other applications can also be found in macroeconomics (Hodrick and E. Prescott, 1997),
climate research (Baillie and Chung, 2002) and social sciences (Levitt, 2004).
Various trend filtering methods have been
developed to recover the underlying mean vector from noisy data. 
We refer  Kim {\it et al.} (2009)  for a complete review of
many different trend filtering methods and  applications.

\subsection{Model assumptions and some background}
\begin{figure}[tp]
\centering
 $$\scalebox{0.5}[0.5]{\includegraphics{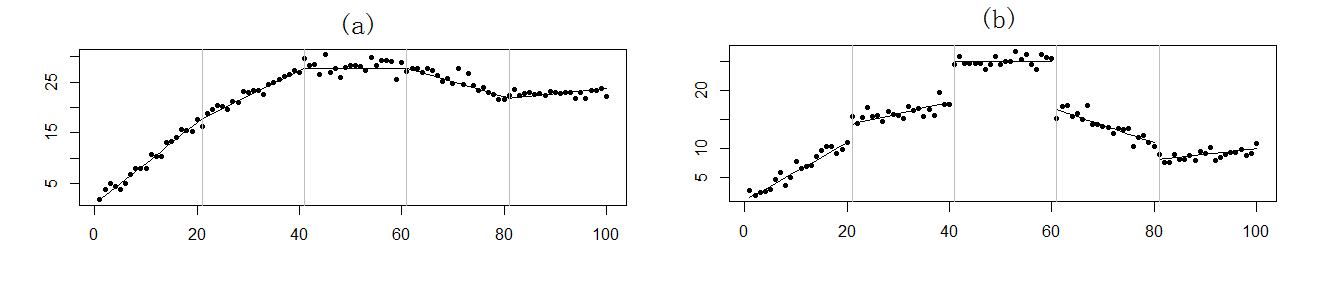}}$$
  \caption{Linear Trend toy example. Dots represent observations at some locations. The
  underlying mean trend are connected by solid lines.
  (a): piece-wise linear trend is jointed; (b): piece-wise linear trend is not jointed}\label{fig1}
\end{figure}

Consider a model
\bel{linear model}
{\rm (I)} \quad y_t&=\mu^0_t+\veps_t, ~1\le t\le n,
\eel
where $\veps_t$ is the random noise
with mean $0$ and variance $\sigma^2$, $1\le t\le n$.
The interest is to recover the mean vector
with joint piece-wise linear trend, meaning the underlying mean vector in model \eqref{linear model}
satisfying:
\bel{piece-linear}
{\rm (\II)}\quad \mu_t^0&=a_j+b_j t, ~ t_{j-1}\le t\le t_{j}-1, ~j=1,\cdots, J+1,~1\le t\le n,
\eel
and
\bel{joint-linear}
{\rm (\III)}\quad  a_{j-1}+b_{j-1} t_j=a_{j}+b_{j} t_j, ~   1\le j\le J.
\eel
Here $t_0=1$ and  $t_{J+1}=n+1$. For $1\le j\le J$, $t_{j}\in\{1,\cdots,n\}$ denote
change points or kink points
where  consistent linear trend changes.
The $(a_j, b_j), j=1,\cdots, J+1$ are $J+1$ pairs of
local intercepts and slopes.
Model assumption (\III) requires that true means at the kink point to be fitted consistently from
 two corresponding adjacent linear trends.  See a toy example in  Figure \ref{fig1} (a).
To recover the mean vector $\bmu^0$ under model I--\III,
one can always get a maximum likelihood or least squares estimation
of those $(a_j, b_j)$'s first by controlling
the number of the kink points.
To list a few, one can see for example Feder (1975a,b), Bai and Perron
(1998) and Bhattacharya (1994).

 However, such an dynamic optimization
approach is computational expensive (Hawkins, 2001).
Since the introduction of the well-known least absolute shrinkage estimator (LASSO) in
 Tibshirani (1996), the $\ell_1$ regularization technique has been widely used in many problems
when the underlying model or the true coefficients vector has some sparse properties.
Here sparsity mean the true model containing many zeros coefficients, but only a few non-zero ones.
The change point detection problem can be treated as  a typical high-dimensional sparse model
in terms of two properties:
the dimension is high since the number of
unknown means equals $n$, the model is sparse since there are
 and only a few true change points.
 Thus an $\ell_1$ regularization approach can be applied to detect non-zero changes,
 and therefore identify the change points.
For example, when the mean vector in \eqref{linear model} is piece-wise constant,
the jump (adjacent difference mean) vector consists of
most  zeros except only a few non-zeros where
 abrupt changes occur. One can
penalize the $\ell_1$ norm of the jump vector to obtain
a piece-wise constant mean estimation,
\bel{model-fusion}
\hbmu_{\rm TV} (\lm_n)=\argmin\{(1/2)\sum_{t=1}^n (y_t-\mu_t)^2+\lm_n\sum_{t=2}^n |\mu_t-\mu_{t-1}|\},
\eel
where $\lm_n\sum_{t=2}^n |\mu_t-\mu_{t-1}|$ for some positive $\lm_n$ is
a total variation (TV) penalty.
Both theoretical and computational properties of $\hbmu_{\rm TV} (\lm_n)$
in \eqref{model-fusion}
have been well studied by Harchaoui and L\'{e}vy-Leduc (2010) and Rinaldo (2009).

Model in \eqref{model-fusion}
was used to detect abrupt change points where
all $b_j$'s are $0$ in linear model \eqref{linear model}.
A similar idea can be adopted for the  recovery of joint piece-wise linear trend.
For $3\le t\le n$, we denote
$\beta_t=\mu_{t}+\mu_{t-2}-2\mu_{t-1}$ as the
potential slope changes at $t-1$ and
$\bbeta=(\beta_1, \cdots, \beta_n)'$. Since
there are only a few local slopes for the underlying slope change vector,
$\bbeta^0$ exhibits some sparse property.
Thus, instead of controlling the number of non-zero slope changes  directly,
 one can obtain
 a piece-wise linear mean trend estimation for model I--\III~
by penalizing  the $\ell_1$ norm of the slope change vector,
\bel{linear filtering}
\hbmu(\lm_n)=\argmin\left\{(1/2)\sum_{t=1}^n (y_t-\mu_t)^2+\lm_n\sum_{t=3}^{n} |\mu_{t}+\mu_{t-2}-2\mu_{t-1}|\right\},
\eel
where $\lm_n>0$ is a tuning parameter controlling the number of estimated linear pieces.
Larger $\lm_n$ will generate smaller number of joint linear pieces.
Model \eqref{linear filtering} is the  $\ell_1$ linear filtering  method
studied in  Kim {\it et al.} (2009),
as a comparison
with the  Hodrick-Prescott (H-P) filtering (Hodrick and Prescott, 1997),
a trend filtering approach to recover the piece-wise quadratic curve.
 Kim {\it et al.} (2009) discussed some basic properties of $\hbmu(\lm_n)$:
\begin{itemize}
 \item [] {\it (P1): As $\lm_n\to 0$, then $\hbmu(\lm_n)$ converges to $\by=(y_1,\cdots,y_n)'$.}
\item [] {\it (P2): As $\lm_n\to \infty$, $\hbmu(\lm_n)$ converges to  the best affine fit to $\by$.}
\end{itemize}
Tibshirani and Taylor (2011)
 also provided some interesting dual algorithms to solve solution paths of
a linear trend filter (LTF) in \eqref{linear filtering}.

However,  many important questions on the properties of an LTF have not been answered.
Will an LTF  \eqref{linear filtering} find
all kink points  asymptotically? If all kink points are detected, will
the joint piece-wise linear mean trend be recovered consistently?
How to choose an optimal $\lm$ to obtain a well-behaved LTF?
In this paper, we will investigate some asymptotic properties of
an LTF under some sufficient conditions.
Specifically, under the joint piece-wise linearity assumption in model I--\III,
 we  will first investigate some rate estimation consistency of
 an LTF if a tuning parameter is well chosen.
Then, we will
provide some sufficient conditions under which
all underlying multiple kink points can be detected correctly in a large probability.
More importantly, those slope changes in direction can be recovered consistently for a well chosen $\lm$.
As a by-product, we will justify that
a weak irrepresentable condition is not needed for the consistent change point detection.


\subsection{Notations and Preliminaries}\label{sec-notation}
We  list some preliminaries and notations to end this section.
Suppose there are $J+1$  linear segments, separated by
kink points $t_j$'s for $1\le j\le J$. We make the following notations:
\begin{itemize}
 \item  $\cJ=\{t_{j}, 1\le j\le J\}$, the collection of the  kink points;
 \item  $\cB_j=\{t_{j-1},\cdots,t_j-1\}$, the subset of $j$th segment;
 \item $|\cB_j|$, the cardinal value of $\cB_j$;
\item Suppose $b_1, \cdots, b_J, b_{J+1}$ are all local slopes. Then the sub-differentials
 \bel{eq-ck}c_j=\partial\sum_{k=2}^{J+1} |b_k-b_{k-1}|/\partial b_j=
\left\{
\begin{array}{ll}
\sgn(b_j-b_{j-1})-\sgn(b_{j+1}-b_j) &\quad {\rm for}~3\le j\le J\\
\sgn(b_{J-1}-b_{J-2})& \quad {\rm for}~j=J+1\\
-\sgn(b_2-b_1) & \quad {\rm for}~j=2
\end{array}
\right.,
\eel
where $\sgn(x)=1,s,-1$ if $x>0, =0, <0$ with $-1<s<1$;
\item  Correspondingly, ${\cJ^0}=\{t^0_j, 1\le j\le J^0\}$, the true kink point set; $b_j^0, 1\le j\le J^0+1$ are the underlying
local slopes
 $\cB_j^0=\{t_{j-1}^0,\cdots,t_j^0-1\}$ is the true $j$th segment and $|\cB_j^0|$ is its
cardinal value;
\item $b_{\min}^0=\min_{1\le j\le J^0} |\cB_j^0|$, the smallest segment size among all  linear pieces;
\item $a_n=\min_{1\le j\le J^0} |\mu_{t_j+1}^0+\mu_{t_j-1}^0-2\mu_{t_j}^0|=\min_{1\le j\le J^0} |b_{j+1}^0-b_{j}^0|$ for $j\in \cJ^0$, the smallest slope change at true kink points.
\end{itemize}
If $\hbmu(\lm_n)$ is an LTF  for $\lm_n>0$, then 
 $\widehat t_{j+1}(\lm_n)$, $\widehat\cJ(\lm_n)$, $\widehat\cB_j$ are defined correspondingly.
We sometimes omit $\lm_n$ from the estimation without causing any confusion.

\subsection{Structure of the paper}
The rest of the paper is presented as follows. In Section 2, we give two different transformations
of the $\ell_1$ linear trend filtering model and discuss some
 corresponding computational and analytical  properties of two types of
 LTF solutions.
 We present our main asymptotic results in Section 3. In this section,
 we provide some sufficient conditions under which an LTF can have some
 rate estimation consistency and detect those kink points consistently.
The effect of weak irrepresentable condition on change point detection
 is also discussed in this section.
In Section 4, we provide some numerical studies containing both simulation studies.
 We summarize the paper with some discussions in Section 5.
 Finally, we give all technical proofs in the Appendix.

\section{Analytical properties of $\ell_1$ trend filter}\label{sec-ltf}
Let the jump value, $\nu_{t}\equiv\mu_{t}-\mu_{t-1}$ be
the slope  between $t-1$ and $t$ for $2\le t\le n$.
Then the slope change $\beta_t
   = \mu_{t}+\mu_{t-2}-2\mu_{t-1}=\nu_{t}-\nu_{t-1}$, for $3\le t\le n$.
   To unify the notation, we also let $\nu_1=\mu_1$,
 $\beta_1=\mu_1$, and $\beta_2=\mu_2-\mu_1$.
 Below we give two different expressions
 of the linear trend filtering model in \eqref{linear filtering}.

 \subsection{Total variation transformation}\label{sec-tv}
Under model assumptions I--\III, the underlying slope vector
 $\bnu^0=(\nu_1^0,\cdots,\nu_n^0)' $ is piece-wise constant
with only a few  abrupt changes.
Thus, we can rewrite model \eqref{linear filtering}
into a
penalized regression model of slope vector $\bnu=(\nu_1,\cdots, \nu_n)'$
with the total variation penalty,
\bel{total variation}
\begin{array}{ll}
\hbnu(\lm_n)&=\argmin f(\bnu,\lm_n)\\
&=\argmin\left\{(1/2)\sum_{t=1}^n (y_t-\sum_{j=1}^n x_{tj}\nu_j)^2+\lm_n\sum_{t=3}^{n} |\nu_{t}-\nu_{t-1}|\right\},
\end{array}
\eel
where $x_{tj}=1$ for $1\le j\le t$ and $0$ otherwise for $1\le t\le n$.
Thus, by modifying the pathwise decent algorithm in Friedman {\it et al.} (2007),
 we can find  an estimation of the slope vector  $\hbnu$ first,
  and then use the cumulative sum to obtain an estimation of $\hbmu$. Below we
give a detailed description of the modified pathwise decent algorithm.
\\
{\bf Modified Pathwise Decent Algorithm}
\begin{itemize}
\item[] 1. Start from $\lm=0$.
\item[] 2. Increase $\lm$ with a reasonable small value and run
the following {\it decent} step and {\it fusion} step, until no further changes occur.
\item[] 3. Repeat 2 until a target $\lm$ is reached.
\end{itemize}
Suppose  $\widetilde\bnu$ is the slope vector obtained from the last step. Then
the current step includes both {\it decent} cycle and {\it fusion} cycle  as follows.
\begin{itemize}
\item [] {\it Decent cycle:}
In \eqref{total variation}, check $\partial f/\partial \nu_k=0$
 for  $\nu_k$ belonging to the following three intervals: $$(-\infty, \min\{\tnu_{k-1}, \tnu_{k+1}\}],~
( \min\{\tnu_{k-1}, \tnu_{k+1}\},\max\{\tnu_{k-1}, \tnu_{k+1}\}],~(\max\{\tnu_{k-1}, \tnu_{k+1}\}, \infty),$$
where $\tnu_{k-1}$ and $\tnu_{k+1}$ are solutions from the last step.
If no solution is found, update $\nu_k$ into the one between  $\tnu_{k-1}$ and $\tnu_{k+1}$
such that  $f$ decreases more.
Specifically, we
 solve
\bel{eq:decent}
\nu_k=(n-k+1)^{-1}\left(\sum_{i=k}^n y_i -\sum_{i=1}^n a_{ik}\tnu_i+\lm\cdot g\right),
\eel
 where $a_{ik}=n-k+1, 0$ and $n-i+1$ for $i<k$, $i=k$ and $i >k$,
where $g=0,\pm 1$ and $\pm2$ for $k=1$, $2$ and $3$ for $3\le k\le n-1$. Here
 ``$+$'' or ``$-$''
is decided in terms of which interval the $\nu_k$ is checked. For example,
if
$\tnu_{k-1}<\tnu_{k+1}$ for some $3\le k\le n-1$, then $g=2$ and $-2$ for
$\nu_k<\tnu_{k-1}$ and $\nu_k>\tnu_{k+1}$, respectively.
\item []  {\it Fusion cycle:}
Enforcing $\nu_k=\nu_{k-1}=\cdots=\nu_{k-m}$ for $1\le m\le k-1$
and assuming $\nu_k=\cdots=\nu_{k-m}=\alpha$
 in the penalized objective function $f$,
check $\partial f/\partial \alpha=0$ for $\alpha$ belongs to any of the following three intervals:
\bel{eq-interval of alpha}
(-\infty, \min\{\tnu_{k-m}, \tnu_{k+1}\}],~
( \min\{\tnu_{k-m}, \tnu_{k+1}\},\max\{\tnu_{k-m}, \tnu_{k+1}\}], ~ (\max\{\tnu_{k-m}, \tnu_{k+1}\}, \infty).
\eel
If a solution $\alpha$ can be found, then we accept the fusion setting.
To be more specific,  we  $\alpha$ is
\bel{eq:fusion}
\left(\sum_{i=k-m}^n y_i d_{m+i+1-k} -
\sum_{i=1}^{m+n+1-k} d_i \sum_{i=1}^{k-m+1} \tnu_i
-\sum_{i=k+1}^{n} ( \tnu_i\sum_{j=m+1}^{m+1+n-i} d_j ) +\lm g\right)
\left/\sum_{i=1}^{m+n-k+1} d_i^2\right.,
\eel
for three intervals in \eqref{eq-interval of alpha},
where $d_i=i$ and $m+1$ for $1\le i\le m$ and $i\le m+1$, and $g$ is defined in the decent step.
\end{itemize}
Here (\ref{eq:decent}) and (\ref{eq:fusion}) can be derived by direct computation.

\subsection{LASSO transformation}\label{sec-lasso}
Another possible approach to obtain $\hbmu(\lm_n)$ is to
consider model  \eqref{linear filtering} as
a LASSO model of the slope change vector $\bbeta=(\beta_1,\cdots, \beta_n)'$,
\bel{lasso}
\hbbeta(\lm)=\argmin\left\{(1/2)\sum_{t=1}^n (y_t-\sum_{j=1}^n z_{tj}\beta_j)^2+\lm_n\sum_{j=3}^{n} |\beta_{j}|\right\},
\eel
where $z_{t1}=1$ for $1\le t\le n$, $z_{tj}=t-j+1$ for $j\le t$ and $z_{tj}=0$ for $j>t$.
Thus the existing algorithm for LASSO can be adopted
to solve $\bbeta=(\beta_1,\cdots,\beta_n)'$ in \eqref{lasso} first.
A mean estimation can be obtained by $\hbmu(\lm)=\bZ\hbbeta(\lm)$,
where $\bZ$ is matrix consisting of all $z_{tj}$'s.
Theoretically, for a given tuning parameter $\lm$, both
\eqref{total variation} and \eqref{lasso} should provide
the same solution.
However, since the tuning parameter selection technique is involved,
those two approaches can provide different final trend filters.
Combining with some existing tuning parameter selection techniques,
LASSO model \eqref{lasso} turns to
 generate more non-zero $\hbeta_j$'s with small values around the true
kink points than the pathwise algorithm does.
 In practice, the pathwise algorithm is preferred if one is more interested
 recovering those change points. However, LASSO model
is preferred is one cares more about the mean trend estimation.
In section \ref{sec-sim}, we use some simulation studies to demonstrate those differences in more details.
We also provide a theoretical justification in Section \ref{sec-sign}.

In the next section, we investigate some asymptotic properties of an LTF,
$\hbmu(\lm_n)$ in  \eqref{linear filtering}. 
We provide some sufficient conditions under which a well-behaved
$\hbmu(\lm_n)$ can be reached.

\section{Asymptotic properties}\label{sec-theory}
In this section, we study the asymptotic properties of an LTF $\hbmu_n$.
In some cases, we aim  to find
 an almost ``unbiased'' estimator of the mean trend vector, which motivate us
to obtain some rate estimation consistency properties under some conditions in Section \ref{sec-est}.
In other cases, we are more interested in the recovery of  kink points,
which motivate us to investigate  some sufficient conditions under which
those underlying kink points can be identified consistently.

 We first make the following assumption on the random noise.

{\it (A1). Random noise $\veps_i$'s  are i.i.d. with mean $0$ and finite variance $\sigma^2$.
Furthermore, they are sub-Gaussian in the sense that $E[\exp(t\veps_i)]\le \exp(\sigma^2 t^2/2), 1\le i\le n.$}

\subsection{Estimation consistency properties}\label{sec-est}
Consider a multiple change point model in  I--\III.
In order to obtain an almost ``unbiased'' estimator of the   unknown mean vector $\bmu^0$
from model \eqref{linear filtering}, 
 we make the following additional assumption on the underlying $\bmu^0$.

 {\it (A2). The underlying mean  $\bmu^0$ in model \eqref{linear model} has at most $J_{\max}$ local linear pieces.}

From (P2) in Section \ref{sec-ltf}, it is always reasonable
to generate an LTF $\hbmu(\lm_n)$ with finite number of linear pieces.
\begin{lemma}\label{lem:lm}
There exists $\lm_0>0$ such that
$\hbmu(\lm_n)$ always has at most $J_{\max}$ local linear pieces for any $\lm>\lm_0$.
\end{lemma}
The proof of Lemma~\ref{lem:lm} is skipped since it is a direct result of (P2) in Section 1.
Then we have the following  rate consistency of the mean estimation.
\begin{theorem}\label{thm-est}
Suppose (A1) and (A2) hold for model I--\III.
Let  $$\lm_n=16^{-1}\sigma A (BJ_{\max}\log n)^{1/2} n^{-1/2} - 32^{-1/2}\sigma(J_{\max})^{1/2} n^{-1/2}$$
 for some $0<A<1$ and $B>2(1-A)^{-2}$. Suppose $\lm_n>\lm_0$ for the $\lm_0$ in
Lemma \ref{lem:lm}.
Then for an LTF  $\hbmu(\lm_n)$ in \eqref{linear filtering},
$$
P\left(n^{-1/2}\| \hbmu-\bmu^0\|_2 \ge \sigma(BJ_{\max}\log n/n)^{1/2}\right)\le J_{\max} n^{[1-B(1-A)^2/2]J_{\max}}.
$$
\end{theorem}
Theorem \ref{thm-est} reveals that we can obtain a consistent estimator of the underlying
mean vector $\bmu^0$ by
choosing  $\lm_n$ appropriately. In addition, the consistency rate is
$O(\log(n)/n)^{1/2})$, which is same as the one obtained by Harchaoui and L\'{e}vy-Leduc (2010)
for the piece-wise constant model in model \eqref{model-fusion}.
Such a consistency rate, $O(\log n/n)^{1/2}$, is also comparable with the optimal rate  obtained by Yao and Au (1988), $O(n^{-1/2})$.
In their work,  the least squares estimation method is used to recover
the piece-wise constant when the number of change-points is bounded.
  We postpone the detailed proof to the Appendix.

\subsection{Sign consistency properties}\label{sec-sign}
In the last section, we provide some rate estimation consistency of the mean estimator $\hbmu(\lm_n)$ if
the number of kink points for  both $\hbmu(\lm_n)$ and $\hbmu^0$ are bounded. In many real applications, we
are more interested in the detection of underlying kink points  where the underlying linear trends change.
In this section, we investigate the consistency of the kink points detection.  More specifically,
we provide some sufficient conditions under which  not only the locations but also the directions of
those slope changes  are recovered with a large probability.
We  make the following  assumptions on the underlying
model and the tuning parameter $\lm_n$:

{\it (B1).  $ \lm_n\le 2 a_n b_{\min}^0$;}

{\it (B2). $a_n^2 (b_{\min}^0)^3\to \infty$;}

{\it (B3). (a) $(\lm_n)^{1/2}\to \infty$; (b) $(2+\log((n-J^0))/\lm_n <28\delta/\sigma$ for some constant $0<\delta<1$.}

Here (B2) requires either $a_n$, the smallest slope changes between any two adjacent linear segments,
or  $b_{\min}^0$, the smallest linear segment size  to be large enough.
Assumptions in (B3) require the tuning parameter $\lm_n$ to grow with $n$. In addition,
 (B3-b) also provides  a lower bound of the growth rate.
(B1) provides some further information on the growth speeds among $\lm_n$,
$a_n$ and $b_{\min}^0$. Notice that there is some redundancy among those conditions. For example,
 one can also deduce (B2) from (B1) and (B3-a). Here we list all those conditions for  a better
understanding of the growth rates of  $a_n$, $b_{\min}^0$ and $\lm_n$.

Denote an event for identifying all kink points
correctly as,
$$\cS^{1n}(\lm_n)=\{\widehat\cJ(\lm_n)=\cJ^0\}.$$
Furthermore, a stronger
event  for detecting directions of slope changes
correctly is,
$$\cS^{n}(\lm_n)=\left\{\cS^{1n}(\lm_n)
 \bigcap \{ \sgn (\hmu_{i+1}(\lm_n)+\hmu_{i-1}(\lm_n)-2\hmu_{i}(\lm_n))=\sgn (\mu^0_{i+1}+\mu^0_{i-1}-2\mu^0_{i}), \forall i\in \cJ^0\} \right\}.$$
 Here $\cS^{n}(\lm_n)$ is stronger than $\cS^{1n}(\lm_n)$ since
 not only  locations of  all kink points (where $\mu^0_{i+1}+\mu^0_{i-1}-2\mu^0_{i}\neq 0$) are detected, but also
  directions of slope changes (sign of $\mu^0_{i+1}+\mu^0_{i-1}-2\mu^0_{i}$) at those locations are correctly recovered.
We call a kink point having a positive (negative) sign when slope increases (decreases) at this location.
Analogous to the variable selection,  we give the following two definitions on the selection
consistency and sign consistency of kink points.
\begin{definition}
$\hbmu$ is kink point detection consistent if $P(\cS^{1n}(\lm_n))\to 1$ when $n\to \infty$.
\end {definition}
\begin{definition}
$\hbmu$ is kink point sign consistent if $P(\cS^{n}(\lm_n))\to 1$ when $n\to \infty$.
\end{definition}
In general, it is much more complicated
to check the event $\cS^{1n}$ directly.  So
we investigate the kink point sign consistency in Definition 2
instead of the kink point detection consistency in Definition 1.
In the following theorem, we provide some sufficient conditions under which
an LTF $\hbmu(\lm_n)$ is kink point sign consistent.
\begin{theorem}\label{thm-sign}
Suppose (A1) and (B1-B3) hold  for a time series model I--\III.
Then for an $\ell_1$ trend filter $\hbmu(\lm_n)$ in \eqref{linear filtering}, we have
$$
\lim_{n\to \infty} P(\cS^{n}(\lm_n))=1.
$$
\end{theorem}
 The proof of Theorem~\ref{thm-sign} is given in the Appendix.
Theorem \ref{thm-sign} indicates that model
\eqref{linear filtering} can recover all kink points
of  a joint piece-wise linear mean trend. Furthermore, the LTF based on a well chosen $\lm_n$ can catch all
 directions of slope changes at those kink points with a large probability.
 This result is the extension of the study  in Rinaldo (2009),
  where change point locations are recovered for the  piece-wise constant trend
  using the total variation penalty.

\subsection{Additional comments} \label{sec-com}
We make two additional comments regarding the above asymptotic results in Section \ref{sec-est} and \ref{sec-sign}.

{\it Comment 1: Weak irrpresentable condition  is not necessary
 for kink point sign consistency.}

In Section \ref{sec-lasso}, we discussed that the $\ell_1$ trend filtering model can be also written into
a LASSO model in \eqref{lasso} with the low triangular design matrix $\bZ$.
Theorem 2 of Zhao and Yu (2006) states that
that the weak irrepresentable condition
is a necessary condition for
a LASSO solution to be sign consistent under two regularity conditions.
Consider a general LASSO model,
\bel{general lasso}
\hbbeta^{(g)}(\lm)=\argmin\left\{(1/2)\sum_{i=1}^n (y_i^{(g)}-\sum_{j=1}^p x^{(g)}_{ij}\beta^{(g)}_j)^2 +\lm \sum_{j=1}^p|\beta_j^{(g)}|\right\},
\eel
where
$(y_i^{(g)}, x^{(g)}_{i1},\cdots,x^{(g)}_{ip})$ and
 $(\beta_1^{(g)},\cdots,\beta_p^{(g)})' $ represent the observed data and
 coefficients vector in the general regression model.
Let $\bX^{(g)}=(\bX^{(g)}_{\bone},\bX^{(g)}_{\btwo})$ be the covariate matrix,
where $\bX^{(g)}_{\bone}$ and $\bX^{(g)}_{\btwo}$ include only the important and unimportant covariates, respectively.
Let $\bs^{(g)}_{\bone}=\sgn(\bbeta_{\bone})$ consist of sign mappings of  non-zero coefficients in the true model.
Then  model \eqref{general lasso} satisfies the  {\it weak irrepresentable condition} if
\bel{wic}
|\bX^{(g)'}_{\btwo}\bX^{(g)}_{\bone}(\bX^{(g)'}_{\bone}\bX^{(g)}_{\bone})^{-1}\bs^{(g)}_{\bone}|<\bone.
\eel
\begin{lemma}\label{lem:wic}
(Zhao and Yu, 2006) Supose two regularity conditions
are satisfied for the designed matrix $\bX^{(g)}$:
(1) there exists a positive definite matrix $C$ such that
the covariance matrix $(\bX^{(g)})'\bX^{(g)}/n\to C$ as $n\to \infty$, and
(2) $\max_{1\le i\le n}(\bx_i^{(g)})'\bx_i^{(g)})/n \to 0$ as $n\to \infty.$
Then LASSO is general sign
consistent only if there exists $N$ so that the weak irrepresentable condition in \eqref{wic} holds for $n > N$.
\end{lemma}
Unfortunately, the LASSO model in \eqref{lasso}
does not satisfies the
weak irrepresentable condition (See an counter example in the Appendix).
However, there is no contradiction between the sign consistency result in Theorem  \ref{thm-sign}
and Lemma \ref{lem:wic}
since those two regularity conditions in
Theorem 2 in Lemma \ref{lem:wic}  are not satisfied for the design matrix $\bZ$
in \eqref{lasso} because of the following lemma.
\begin{lemma}\label{lem:reg1}
The design matrix $\bZ$ in model \eqref{lasso}  has two properties:
(a) $\rho_1<1/(4n)\to 0$ when $n\to \infty$,
where $\rho_1$ is the smallest
eigenvalue of $\bZ'\bZ/n$ and
(b)  $\max_{1\le i\le n} \bz_i'\bz_i/n \ge n^2/4$.
\end{lemma}
 Lemma \ref{lem:reg1} can be verified easily and
we skip the detailed proof in this manuscript.
Two properties (a) and (b) in Lemma \ref{lem:reg1} means
the both regularity conditions (1) and (2) in Lemma \ref{lem:wic}
are violated.  Thus Theorem \ref{thm-sign} is
not against Lemma \ref{lem:wic}. In other words,
Theorem \ref{thm-sign} indicates that
the weak irrepresentable condition is not necessary
for the change point detection.

{\it Comment 2: An LTF may not reach the estimation consistency and sign consistency simultaneously.}

The rate estimation consistency in Theorem \ref{thm-est} holds
for $\lm_n=O(\log (n)/n)$. However,
from (B3-b), we know one of the sufficient conditions for the sign consistency in
 Theorem \ref{thm-sign} requiring  $\lm_n>O(\log (n))$.
So an $\ell_1$ trend filter may not
be able to reach the estimation consistency and sign consistency simultaneously.
However, this claim is not theoretically justified since
 all conditions assumed in both Theorem \ref{thm-est}
and \ref{thm-sign} are sufficient.



%
%
%
%
%
%
%
\section{Numerical studies}\label{sec-num}
In this section, we use some simulation studies to demonstrate the performance of the $\ell_1$ trend filter.
\subsection{Tuning parameter selection}
As stated in Section \ref{sec-ltf}, for every $\lm>0$,
we can first find an optimizer  $\hbnu(\lm)$  of \eqref{total variation} using the modified
pathwise decent algorithm. For such a $\lm>0$, an $\ell_1$ trend filter
is obtained from corresponding cummalative summation.
Since $\lm$ controls  the  number of abrupt changes in $\hbnu(\lm)$,
it is important to choose  an optimal tuning parameter, $\lm_{\rm opt}$,
from a sequence of $\lm\in(\lm_{\min}, \lm_{\max})$, where $\lm_{\max}$ is
a sufficient large $\lm$ such that
 $\hbmu(\lm_n)$ reaches the best affine fit to $\by$. For a fixed
 $\lm>0$, model \eqref{total variation} is a modeling procedure including both
model selection and model fitting. Zou {\it et al.} (2007) justified the number of
non-zero estimates of a LASSO estimates is an unbiased estimates of the degrees of
the freedom of the LASSO modeling procedure. In addition,
Tibshirani and Taylor (2011) also confirmed
$\widehat k(\lm)+2$ to be an unbiased estimates of the $\ell_1$ trend filtering procedure in
\eqref{linear filtering}, where $\widehat k(\lm)=|\widehat\cJ(\lm)|$ is the
number of estimated kink points. Thus, one can apply different
model selection criteria to choose an optimal tuning parameter $\lm$.
For example, one can adopt the
 Schwarz Information Criterion (SIC) (Schwardz, 1978) to choose the optimal
tuning parameter $\lm_{\rm S}$ as,
 \bel{sic}
\lm_{\rm S}=\argmin\left\{\log\left(\sum_{t=1}^n(y_t-\widehat y_t)^2/n\right)+(k(\lm)+2) \log(n)/n\right\}.
\eel
In addition, Ciuperca (2011)
proposed an M-criterion (MC) to determine the number of change-points of
parametric nonlinear multi-response model, where
the joint piece-wise linear model is a particular case.
Specifically for Gaussian error and least squares regression, we
can also choose optimal $\lm$ using MC as,
 \bel{mc}
\lm_{\rm M}=\argmin\left\{\log\left(\sum_{t=1}^n(y_t-\widehat y_t)^2/n\right) + \widehat k(\lm) (\widehat k(\lm)+1)\log n/n\right\}.
\eel
Ciuperca (2011) demonstrated that MC has some advantages over SIC in terms
of change points detection for several linear trend cases. In the next section, we
 use some simulation studies  to demonstrate the performance of an LTF, where
both SIC and MC are adopted in the LTF modeling procedure.

\subsection{Simulation studies}\label{sec-sim}
In the simulation study, we simulate 100 data sets.
Each data set consists of $n$ observations generated from linear model
\eqref{linear model}. The true linear trend $\bmu^0$ consists of
$k$ linear pieces, with kink points set $\cJ^0=\{nr_1+1,nr_2+1,\cdots, nr_{k-1}+1\}$.
The slope vector for all $k$ pieces $\mathbf{b}=(b_1,\cdots,b_k)'$.
The first linear piece has zero intercept, and the rest intercepts
are derived correspondingly such that all the linear pieces are jointed.
 Let the  signal to noise ratio (SNR) be $\sum_{i=1}^n\mu_i^0/(n\sigma)$. We
 simulate the Gaussian white noise
at three different SNR: low-noise for SNR=$10^4$,
moderate-noise when SNR=$400$,
and heavy-noise when SNR=$25$.
Two cases of jointed linear pieces are considered:

\begin{example}\label{example1}
(Symmetric linear trend with a constant segment) $k=3$, $\{r_1,r_{2}\}=\{0.3,0.7\}$ and $\mathbf{b}=(-30,0,30)$.
\end{example}
\begin{example}\label{example2}
(Linear trend with waggled slope changes) $k=5$, $\{r_1,r_{2},r_{3},r_{4}\}=\{0.2,0.4,0.6,0.8\}$ and $\mathbf{b}=\{-6, 40, -5,  35, -3\}$.
\end{example}

\begin{figure}[tp]
\centering
 $$
 \scalebox{0.3}[0.3]{\includegraphics{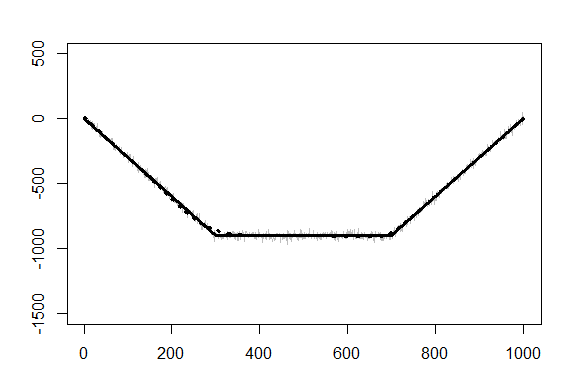}}
 \scalebox{0.3}[0.3]{\includegraphics{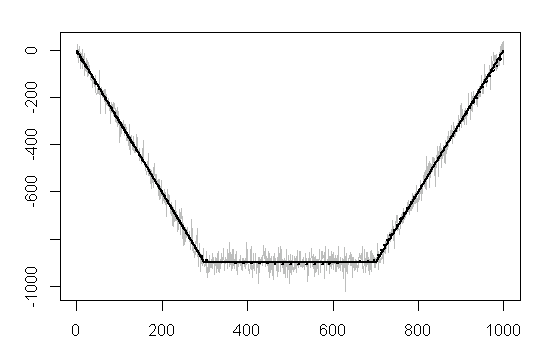}}
 \scalebox{0.3}[0.3]{\includegraphics{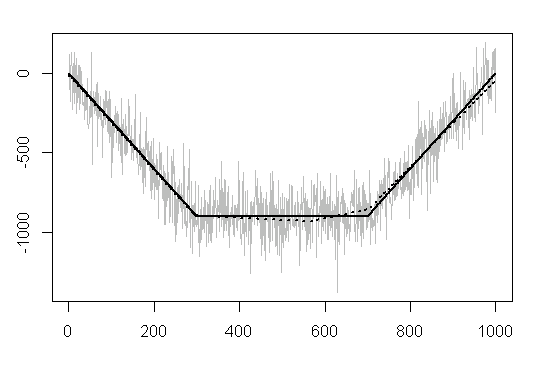}}$$
 $$
 \scalebox{0.3}[0.3]{\includegraphics{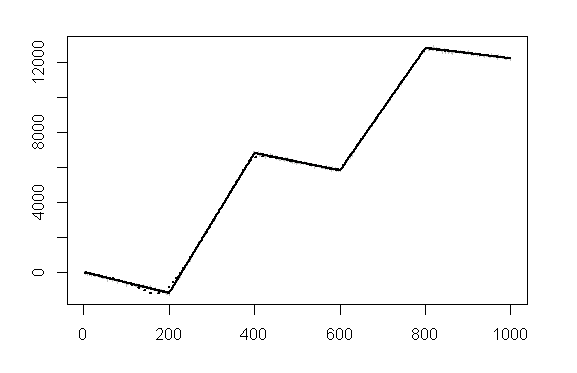}}
 \scalebox{0.3}[0.3]{\includegraphics{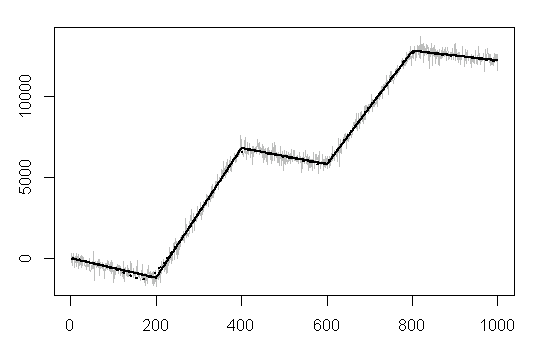}}
 \scalebox{0.3}[0.3]{\includegraphics{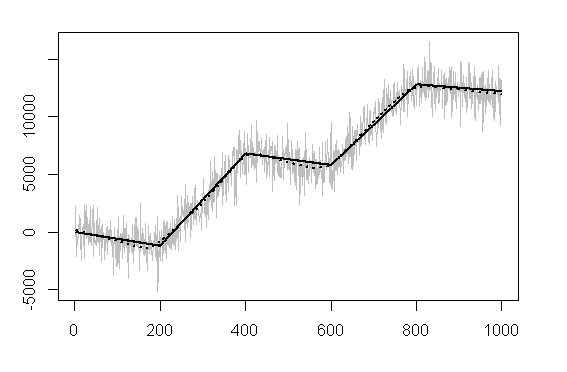}}$$
  \caption{Simulated data examples for all six settings. $n=1000$.
  The solid curve is the underlying  linear trend. The dotted curve
  is the LTF output for $\lm=20a_n b_{\min}^0$.
  The low, moderate and heavy noises are plotted
from the left to the right.
The top and bottom panels are for Example 1 and 2, respectively.}\label{fig2}
\end{figure}

In total there are six different settings in Example \ref{example1} and \ref{example2}.
 For each setting, we simulated 100 data sets with size of  $n=500$ and $1000$.
In Figure \ref{fig2}, we provide different sample data sets for all six settings with
 corresponding true and fitted linear trends.
The low, moderate and heavy noises are plotted
from the left to the right.
The top and bottom panels are for Example 1 and 2, respectively.

We demonstrate the estimation effects by computing the relative error (RE)
as follows:
\bel{eq:lre}
{\rm RE}(\hbmu_n,\bmu^0)=\dfrac{\sum_{i=1}^n(\hmu_i-\mu_i^0)^2}{\sum_{i=1}^n\hmu_i^2}.
\eel
To evaluate the performance of the $\ell_1$ trend filtering model
in terms of the kink points recovery, we compute both
 means and standard deviations of the estimated kink point for all cases.
Similar to Boysen et al. (2009) and Hrchaoui and L\'{e}vy-Leduc (2010), we also
report the Hausorff distance
between $\widehat\cJ$ and $\cJ^0$.
Let $A$ and $B$ are two sets. The Hausdorff distance,
\bel{eq:hd}
{\rm HD}(A,B)=\sup\{\cE(A||B); \cE(B||A)\},
\eel
where $\cE(A||B)=\sup_{b\in B}\inf_{a\in A}|a-b|$.
 We choose the optimal tuning parameter from $\lm\in(0, \lm_{\max})$ using
both  SIC in \eqref{sic} and MC in \eqref{mc}, where
$\lm_{\max}$ is chosen to be the smallest one such as an affine fit is reached.
The simulation results from Example 1 and 2 are summarized in Table 1.

\begin{figure}[tp]
\centering
 $$
 \scalebox{0.3}[0.3]{\includegraphics{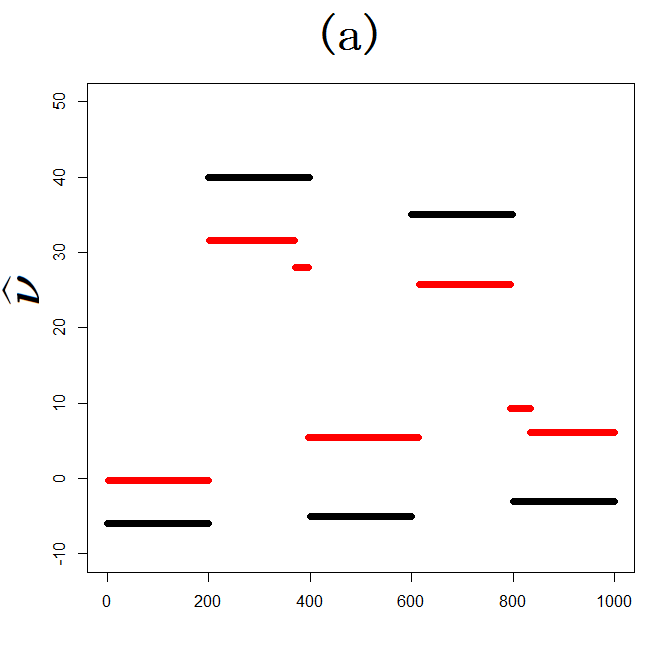}}
 \scalebox{0.3}[0.3]{\includegraphics{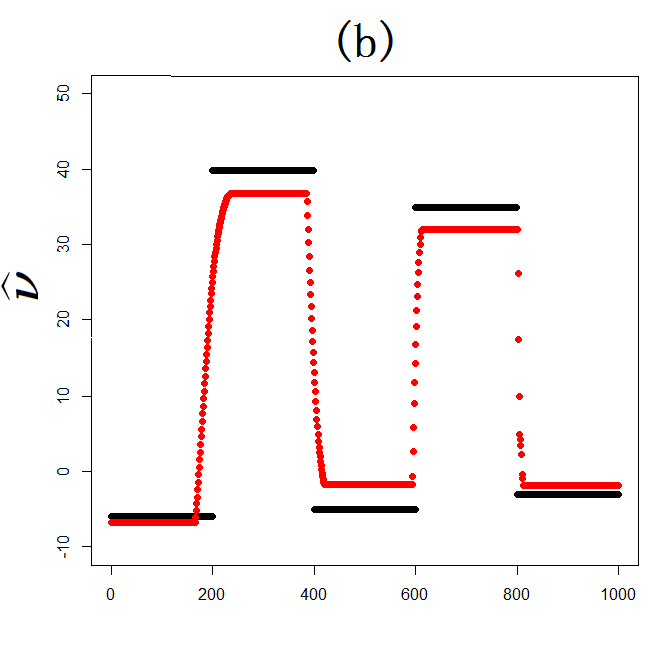}}
$$
  \caption{Fitted slope vector, $\hbnu$,  from a simulated data in Example 2. The left and
right panels are outputs from pathwise decent algorithm
for model \eqref{total variation} and coordinate decent algorithm for \eqref{lasso}, respectively.
  (Observation: grey, True: black, SIC: blue, MC: red).}\label{fig3}
\end{figure}

\begin{figure}[tp]
\centering
 $$
 \scalebox{0.3}[0.3]{\includegraphics{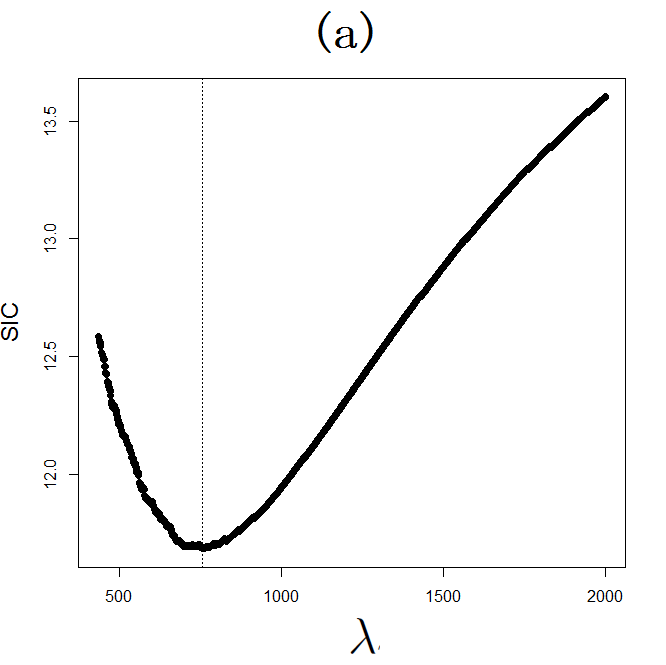}}
 \scalebox{0.3}[0.3]{\includegraphics{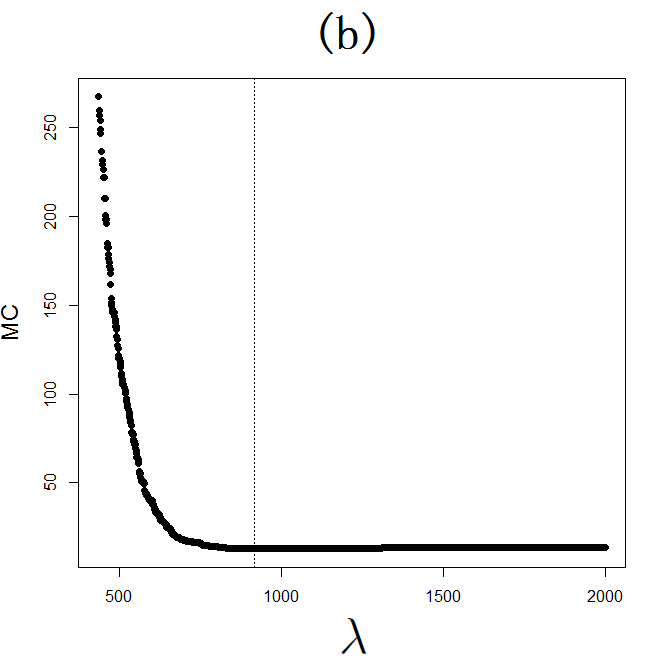}}
$$
  \caption{The SIC and MC curves for a simulated data with $n=1000$ with median noise in Example 2. Left: SIC; Right: MC.
  The vertical lines identify corresponding optimal $\lm$ values}\label{fig4}
\end{figure}

Overall, MC works much better in terms of kink points detection.
However, SIC  generates less bias on mean estimation.
In Figure \ref{fig4} we plot both  MC and SIC curves
for  a simulated data set generated from Example 2.
The behavior of $\hbmu(\lm)$ depends on
$\lm$ tightly such that a larger $\lm$ generates less kink points but trigger
larger estimation biases, while a reasonable $\lm$ for  smaller bias
may not be large enough
for the recovery of underlying kink points.
Such an observation is consistent with Comment 2 in Section \ref{sec-com}.
As a comparison, in  Table 2, 
 we also report some simulation results
 from the LASSO model \eqref{lasso}. In general, the modified pathwise
algorithm performs much better than the LASSO algorithm in terms of the
kink point detection. The MC is used in both approaches.
 In Figure  \ref{fig3}, we further review this phenomenon
by plotting two estimated slope vector $\hbnu$ from
\eqref{total variation} and \eqref{lasso} for
simulated data example in Example 1.
We found that the $\hbnu$ changes abruptly (Figure 3(a)) generated from
the pathwise algorithm and model \eqref{total variation}. However,
 $\hbnu$ from the LASSO algorithm turns to change
gradually around  kink points (Figure 3(b)). 
It tells us that, even though both \eqref{total variation} and \eqref{lasso}
are able to find a $\ell_1$ trend filter, the final solutions can be
very different due to the different effects of the
 tuning parameter selection techniques.


\section{Discussion}
In this paper, we study the asymptotic properties of
 the joint linear trend recovery using the $\ell_1$ regularization approach.
By assuming the true model to be piece-wise linear,
we investigate both the estimation consistency and sign consistency  of an  $\ell_1$ trend filter.
In terms of estimation consistency, the consistency rate
is optimal up to a logarithmic factor  if the dimension of any linear space
where the true model and its estimates belong to is bounded from above.
In terms of sign consistency,  we justify that
an $\ell_1$ trend filter  can not only recover the locations
where  the underlying linear pieces connect
but also distinguish those slope changes in direction
 with high probability under reasonable conditions.
Thus, by choosing the
tuning parameter $\lm$ properly, we can reach a well-behaved
linear trend filter to recover the underlying  linear piece-wise
 from some random noises.
The consistency results in this paper amplify
 the study  in  Harchaoui and L\'{e}vy-Leduc (2010) and Rinaldo (2009), where
 the $\ell_1$ regularization approach is used to
 recover the piece-wise constant for  signal approximation.
As a by-product, we also justify that a weak irrepresentable condition is not necessary for the
change point detection. In addition, we evaluate the performances of two alternative expressions of the
$\ell_1$ trend filtering models in terms of both the total variation penalty and the
LASSO penalty. A modified pathwise algorithm is preferred than the
LASSO if the main focus is the kink points detection.

As in many recent studies for penalized regression, our results are proved for the penalty parameter that satisfy the conditions as stated in the theorems.
It is not clear whether the penalty parameter selected using data-driven procedures satisfies those conditions.
However, our numerical study shows a satisfactory finite-sample performance of the $\ell_1$ trend filter.
Particularly, we note that the tuning parameter selected based on  MC seems much better than  the one from SIC for our simulated data.
Tuning parameter selection
 is an important and challenging problem that requires further investigation, but is beyond the scope of the current paper.
\section{Technical proofs}
In this section, we provide proof of main results in Section \ref{sec-theory}.
For the notation's convenience, we sometimes omit $\lm_n$ without causing any confusion.

\noindent{\bf Proof of Theorem \ref{thm-est}}

\noindent In this proof, we omit $\lm_n$ and let $\hbmu=\hbmu(\lm_n)$ defined in \eqref{model-fusion}.
Recall that $\nu_i=\mu_i-\mu_{i-1}$ for $2\le i\le n$ and $\nu_1=\mu_1$. Then $\nu_i^0$ and
$\hnu_i$ are defined correspondingly. For example, $\hbnu=(\hnu_1, \cdots, \hnu_n)'$
with $\hnu_1=\hmu_1$ and $\hnu_i=\hmu_i-\hmu_{i-1}$ for $2\le i\le n$.
From the definition of $\hbmu$ and $\hbnu$, we have
$$
(1/2)\sum_{i=1}^n \left(y_i-\sum_{j=1}^i \hnu_j\right)^2 +\lm_n \sum_{i=2}^{n-1}|\hnu_{i+1}-\hnu_i|\le
(1/2)\sum_{i=1}^n \left(y_i-\sum_{j=1}^i \nu^0_j\right)^2 +\lm_n \sum_{i=2}^{n-1}|\nu^0_{i+1}-\nu^0_i|.
$$
Then
$$
(1/2)\sum_{i=1}^n \left[\sum_{j=1}^i (\hnu_j-\nu_j^0)\right]^2 \le \lm_n
[ \sum_{i=1}^{n-1} |\nu^0_{i+1}-\nu^0_i| -|\hnu_{i+1}-\hnu_i|]+\sum_{i=1}^n[ \sum_{j=1}^{i}(\hnu_{j}-\nu^0_j)\veps_i].
$$
Recall that
$\bx_i=(x_{1i},\cdots,x_{ni})'$ for $1\le i\le n$ and
 $\bX$ is a lower triangle matrix with $1$ for the non-zero element.  We have
$$
(1/2)(\hbnu-\bnu^0)' \sum_{i=1}^n \bx_i\bx_i' (\hbnu-\bnu^0)\le
               2\lm_n \sum_{i=1}^{n}[|\hnu_{i}-\nu_i^0|]+(\hbnu-\bnu^0)'\sum_{i=1}^n \bx_i\veps_i.
$$
and
$$
(1/2)(\hbnu-\bnu^0)'\bX'\bX (\hbnu-\bnu^0)\le
                2\lm_n n^{1/2}\|\hbnu-\bnu^0\|_2+(\hbnu-\bnu^0)'\bX'\bveps,
$$
where $\bveps=(\veps_1,\cdots,\veps_n)'$.
Denote $\Delta(\bw)=\|X(\bw-\bnu^0)\|_2$ and then
$\Delta(\hbnu)=\|X(\hbnu-\bnu^0)\|_2$.
 Let $G(\bw)=(\sigma\Delta(\bw))^{-1}(\bw-\bnu^0)'\bX'\bveps$ for $\bw\in \cR^n$.
Then we have
$$
(1/2)(\Delta(\hbnu))^2 \le 2\lm_n n^{1/2}\|\hbnu-\bnu^0\|_2+\sigma \Delta(\hbnu) G(\hbnu).
$$
Let $r_1\le r_2\le \cdots\le r_n$ be $n$ eigenvalues of $\bX'\bX$.
Then $ r_1>1/4$. Thus,
$$
\Delta(\hbnu) \le 16\lm_n n^{1/2}+2\sigma G(\hbnu).
$$
So for any $\alpha_n>0$, we have
\bel{eq-Gu1}
P\left(\Delta(\hbnu)\ge \alpha_n\right) \le P\left(G(\hbnu)\ge \alpha_n/(2\sigma)-8\lm_n n^{1/2}/\sigma\right).
\eel
We borrow some notations from
Harchaoui and L\'{e}vy-Leduc (2010).
Consider $\{S_K\}_{1\le K\le J^0}$ to be a
collection of linear spaces where $\hbnu$ may belong, where
$S_K$ is a linear space of $K$ dimension.
In addition, from Borell-TIS inequality (Ledoux and Talagrand, 1991), we have
\bel{eq-max}
P\left(\sup_{\bw\in S_K} G(\bw)\ge E[\sup_{\bw\in S_K} G(\bw)] +c \right ) \le \exp\{-c^2/2\}.
\eel
 There exist an $n\times n$
 orthogonal matrix $\bP$ such that $\bX'\bX=\bP'\bLambda \bP$, where $\bLambda=\diag(r_1,\cdots,r_n)$ is a diagonal matrix.
 Let $\cW$ be a $D$-dimensional linear space where $\bw-\bnu^0$ belongs. Then
 we can write $\bw-\bnu^0=\sum_{j=1}^D \alpha_j \bphi_j={\bf \Phi} \balpha$,
 where
  $\bphi_1,\cdots, \bphi_D$ are the orthogonal basis of $\cW$,
  ${\bf \Phi}=(\bphi_1,\cdots, \bphi_D)'$ with $\bPhi'\bPhi=\bI_D$ and, and
 $\balpha=(\alpha_1,\cdots,\alpha_D)'\in \cR^D$. Define
 $$\bb_{\ba} \bb_{\ba}'=\balpha'\bPhi'\bX'\bX\bPhi\balpha]^{-1/2}\balpha'\bPhi'\bX'\bPhi.$$
From the Cauchy-Schwarz inequality, we have
\bel{eq-supG}
 \begin{array}{ll}
\sigma E[\sup_{\bw\in S_K} G(\bw)] & =E[\sup_{\balpha\in R^D}(\balpha'\bPhi'\bX'\bX\bPhi\balpha)^{-1/2} \balpha'\bPhi'\bX'\bveps]\\
& =E[\sup_{\balpha\in R^D}\bb_{\ba}'\bPhi'\bveps]\\
& \le \sup_{\balpha\in R^D}\{(\bb_{\ba} '\bb_{\ba})^{1/2}  E[ (\bveps'\bPhi\bPhi'\bveps)^{1/2}]\}.
\end{array}
\eel
First,
$$E[\bveps'\bPhi\bPhi'\bveps]=\sigma^2 {\rm tr}(\bPhi\bPhi')=\sigma^2 {\rm tr}(\bPhi'\bPhi)\le D\sigma^2.$$
We will prove that $\sup_{\balpha\in R^D}(\bb_{\ba} '\bb_{\ba})^{1/2}\le 1$ as follows
$$
\begin{array}{ll}
\bb_{\ba} '\bb_{\ba}-1&=(\balpha'{\bf \Phi}'\bX'\bX{\bf \Phi}\balpha)^{-1}(\balpha'\bPhi'\bX'\bPhi\bPhi'\bX\bPhi\balpha)-1\\
                    &=(\balpha'{\bf \Phi}'\bX'\bX{\bf \Phi}\balpha)^{-1}[ \balpha' \bPhi'\bX'(\bPhi\bPhi'  -\bI_n )\bX \bPhi\balpha]
\end{array}
$$
Notice that $\bI_n-\bPhi\bPhi'\ge 0$ is an idempotent semi-definite matrix. Eigenvalues of $\bPhi\bPhi'  -\bI_n$ are only $0$ and $-1$.
Therefore, $\bb_{\ba} '\bb_{\ba}-1\le 0$ and $\bb_{\ba} '\bb_{\ba}^{1/2}\le 1$. Thus from \eqref{eq-supG},
\bel{eq-max}
E[\sup_{\bw\in S_K} G(\bw)] \le D^{1/2}\le (2J_{\max})^{1/2}.
\eel
From \eqref{eq-max}, we have
\bel{eq-d}
d_K&\equiv \alpha_n/(2\sigma)-8\lm_n n^{1/2}/\sigma -E[\sup_{\bw\in S_K} G(\bw)]\\
 &>\alpha_n/(2\sigma)-8\lm_n n^{1/2}/\sigma-(2J_{\max})^{1/2}.
 \eel
We denote $c_0=\alpha_n/(2\sigma)-8\lm_n n^{1/2}/\sigma-(2J_{\max})^{1/2}$. Then
if we choose $\alpha_n$ such that $\alpha_n/(2\sigma)-8\lm_n n^{1/2}/\sigma-(2J_{\max})^{1/2}>0$ is satisfied.
Then there exists
$0<A<1$, such that
$$
A\alpha_n/(2\sigma)=8\lm_n n^{1/2}/\sigma+(2J_{\max})^{1/2}.
$$
If we define
 \bel{eq-c1}
 c_0=(1-A)\alpha_n/\sigma,
 \eel
then $d_K>c_0>0$.
From \eqref{eq-Gu1} and \eqref{eq-c1}, we have
$$
\begin{array}{ll}
P\left(\Delta(\hbnu)\ge \alpha_n\right)& \le \sum_{K=1}^{J_{\max}} n^K P\left(\sup_{\bw\in S_K} G(\bw)\ge  \alpha_n/(2\sigma)-8\lm_n n^{1/2}/\sigma\right)\\
                                &=  \sum_{K=1}^{J_{\max}} n^K P\left(\sup_{\bw\in S_K} G(\bw)\ge E[\sup_{\bw\in S_K} |G(\bw)|] +d_K \right)\\
                                &\le J_{\max} n^{J_{\max}} P\left( \sup_{\bw\in S_K} G(\bw)\ge E[\sup_{\bw\in S_K} |G(\bw)|] +c_0 \right)\\
                                     &\le J_{\max} n^{J_{\max}}\exp\{-c_0^2\}\\
                            &=J_{\max}\exp\{J_{\max}\log n-[(\alpha_n/\sigma)(1-A)]^2\}.
\end{array}
$$
Let $(\alpha_n/(2\sigma))^2=J_{\max} B \log n$ and then  $\alpha_n=2\sigma (J_{\max} B\log n)^{1/2}$. Thus,
we have
$$
\begin{array}{ll}
&P\left(n^{-1/2}\|\hbmu-\bmu^0\|\ge \sigma (J_{\max} B(\log n)/n)^{1/2}\right) \\
 &\quad=P(\Delta(\hbnu)\ge \alpha_n)  \\
 &\quad\le J_{\max}\exp\{J_{\max}\log n-BJ_{\max}(\log n)(1-A)^2\} \\
&\quad=J_{\max}\exp\{J_{\max}(\log n) (1-B(1-A)^2/2)\}\\
&\quad=J_{\max} n^{J_{\max}[1-B(1-A)^2/2]}\to 0~{\rm if}~ B>2/(1-A)^2.
\end{array}
$$
$\Box$

 \noindent{\bf Proof of Theorem \ref{thm-sign}}

 \noindent 
Let $\ba=(a_1, \cdots, a_{J+1})'$ and ${\bf b}=(b_1,\cdots, b_{J+1})'$, where
$(a_{j}, b_{j})$ is the intercept and slope of the $j$th local linear pieces.
 For linear model (\ref{linear model}--\ref{joint-linear}),
we can write the penalized loss function  in \eqref{linear filtering} into
 \bel{eq-opt1}
 f(\ba,{\bf b}; \lm_n)=(1/2)\sum_{j=1}^{J+1}\sum_{k\in \cB_j}(y_k-(a_j+b_j k))^2+\lm_n\sum_{j=2}^{J+1}|b_j-b_{j-1}|.
 \eel
 Suppose $\widehat \ba(\lm_n)$ and $\widehat {\bf b}(\lm_n)$ are
the optimizer of \eqref{eq-opt1} for $\lm_n>0$.
We omitted $\lm_n$ for the rest of the proof without causing any confusion.
Let $\widehat\cB_{j(i)}$ represent the index set of the local linear segment where
$\hmu_i$ stays. Correspondingly, $(\widehat a_{j(i)}, \widehat b_{j(i)})'s$ are  the
local intercept and slopes at  $i$ and $\hmu_i=\widehat a_{j(i)}+\widehat b_{j(i)} i$
for $1\le i\le n$.
From the Karush-Kuhn-Tucker condition of the above optimization problem \eqref{eq-opt1},
$\hbmu$ is  an LTF solution if and only if
\begin{eqnarray}\label{kkt-1}
\left\{
\begin{array}{ll}
\sum_{k\in \widehat\cB_{j(i)}} k[y_k-(\widehat a_{j(i)}+ \widehat b_{j(i)} k)]=\lm_n \widehat c_{j(i)} &\quad {\rm for~}\widehat b_{j(i)}\neq\widehat b_{j(i-1)}\\
\arrowvert \sum_{k\in \widehat\cB_{j(i)}} k[y_k-(\widehat a_{j(i)}+ \widehat b_{j(i)} k)] \arrowvert <4\lm_n &\quad {\rm for~}\widehat b_{j(i)}=\widehat b_{j(i-1)}
\end{array}
\right.,
\end{eqnarray}
where $\widehat b_{j(i)}=\widehat b_{j(i-1)}$ also means $\hmu_i-\hmu_{i-1}=\hmu_{i+1}-\hmu_{i}$. Here
 $\widehat c_{j(i)}$ is an corresponding estimation of $c_j$  in
\eqref{eq-ck}.
Define $\bgamma_{j(i)} \equiv(\sum_{k\in \cB_{j(i)}^0} k, \sum_{k\in \cB_{j(i)}^0} k^2)'$. Consider
\bel{eqn-ab}
\widehat a_{j(i)}=a^0_{j(i)}+
+ (\bgamma_{j(i)}'\bgamma_{j(i)})^{-1}\left(\sum_{k\in \cB_{j(i)}^0} k\right)
\left(\sum_{l\in \cB_{j(i)}^0} l\veps_l -\lm_n c_{j(i)}^0\right) ~{\rm for~} i\in \cJ^0,
\eel
and
\bel{eqn-b} \widehat b_{j(i)} =b^0_{j(i)} +
(\bgamma_{j(i)}'\bgamma_{j(i)})^{-1}\left(\sum_{k\in \cB_{j(i)}^0} k^2\right) \left(\sum_{l\in \cB_{j(i)}^0} l\veps_l -\lm_n c_{j(i)}^0\right) ~{\rm for~} i\in \cJ^0.
\eel
Thus
$\widehat {\bf b}$ satisfying \eqref{eqn-b} and
\bel{eqn-b2}
\widehat b_{j(i)}=\widehat b_{j(i-1)},\quad i\notin \cJ^0
\eel
 is a solution of ${\bf b}$  in \eqref{eq-opt1}.
Therefore, from \eqref{kkt-1}, $\cS^{n}$ holds if and only if  $\widehat {\bf b}$
in (\ref{eqn-b}--\ref{eqn-b2}) satisfies
\begin{eqnarray}\label{kkt-2-1}
\sgn(\widehat b_{j(i)}-\widehat b_{j(i-1)})= \sgn(b^0_{j(i)}- b^0_{j(i-1)}) &\quad {\rm for~} i\in \cJ^0
\end{eqnarray}
and
\begin{eqnarray}\label{kkt-2-2}
\arrowvert \sum_{k\in \widehat B^0_{j(i)}} k[y_k-(\widehat a_{j(i)}+ \widehat b_{j(i)} k)]\arrowvert <2\lm_n &\quad {\rm for~}\widehat b_{j(i)}=\widehat b_{j(i-1)}.
\end{eqnarray}
We now first verify  \eqref{kkt-2-1}.  Notice that \eqref{kkt-2-1} holds if
\begin{eqnarray}\label{kkt-2-1-1}
|(\widehat b_{j(i)}-b^0_{j(i)})-(\widehat b_{j(i-1)}- b^0_{j(i-1)})|< |b^0_{j(i)}- b^0_{j(i-1)}|&\quad {\rm for~} i\in \cJ^0.
\end{eqnarray}
Plug \eqref{eqn-b} into \eqref{kkt-2-1-1}, we get
\begin{equation}\label{kkt-2-1-2}
\begin{array}{ll}
&\vline \left[(\bgamma_{j(i)}'\bgamma_{j(i)})^{-1}\sum_{k\in \cB_{j(i)}^0} k^2
\left(\sum_{l\in \cB_{j(i)}^0} l\veps_l -\lm_n c_{j(i)}^0\right)\right]-\\
&\quad \left[(\bgamma_{j(i-1)}'\bgamma_{j(i-1)})^{-1}\sum_{k\in \cB_{j(i-1)}^0} k^2
\left(\sum_{l\in \cB_{j(i-1)}^0} l\veps_l -\lm_n c_{j(i-1)}^0\right)\right]\vline \\
&\quad\quad< |b^0_{j(i)}- b^0_{j(i-1)}|\quad {\rm for~} i\in \cJ^0.
\end{array}
\end{equation}
 Notice that $a_n=\min_{i\in \cJ^0} |b^0_{j(i)}- b^0_{j(i-1)}|$.
We  expand \eqref{kkt-2-1-2} into different inequalities.
Denote $I_{1}$ as
\begin{equation}\label{kkt-2-1-3}
\left\{ \max_{i\in\cJ^0} \left| (\bgamma_{j(i)}'\bgamma_{j(i)})^{-1}\sum_{k\in \cB_{j(i)}^0} k^2
\sum_{l\in \cB_{j(i)}^0} l\veps_l-(\bgamma_{j(i-1)}'\bgamma_{j(i-1)})^{-1}\sum_{k\in \cB_{j(i-1)}^0} k^2
\sum_{l\in \cB_{j(i-1)}^0} l\veps_l \right|\le a_n/2\right\}.
\end{equation}
Denote $I_{2}=I_{21}\cap I_{22}$ and
\begin{equation}\label{kkt-2-1-4}
\begin{array}{ll}
I_{21}&\equiv \left\{ \max_{i\in\cJ^0} \left| (\bgamma_{j(i)}'\bgamma_{j(i)})^{-1}\sum_{k\in \cB_{j(i)}^0} k^2  c_{j(i)}^0 \right| \le a_n/(2\lm_n)\right\}\\
I_{22}&\equiv \left\{ \max_{i\in\cJ^0} \left| (\bgamma_{j(i-1)}'\bgamma_{j(i-1)})^{-1}\sum_{k\in \cB_{j(i-1)}^0} k^2  c_{j(i-1)}^0 \right| \le a_n/(2\lm_n)\right\}.
\end{array}
\end{equation}
Therefore \eqref{kkt-2-1-2} holds if $I_1$, $I_{21}$ and $I_{22}$ hold.
 If (B1) holds, then
$$
\frac{\lm_n}{a_n}<\sum_{k\in \cB_{j(i)}^0} k
       <\frac{1}{2}\left(\frac{\left(\sum_{k\in \cB_{j(i)}^0} k\right)^2}{\sum_{k\in \cB_{j(i)}^0} k^2}+\sum_{k\in \cB_{j(i)}^0} k^2 \right)
       =   \left( 2(\bgamma_{j(i)}'\bgamma_{j(i)})^{-1}\sum_{k\in \cB_{j(i)}^0} k^2 \right)^{-1} .
$$
Thus
\begin{equation}\label{kkt-2-1-5}
P(I_2^c)\le P(I_{21})+P(I_{22})=0.
\end{equation}
We now consider the event $I_1$.
 Let
 $$\tau_i=(\bgamma_{j(i)}'\bgamma_{j(i)})^{-1}\sum_{k\in \cB_{j(i)}^0} k^2\sum_{l\in \cB_{j(i)}^0} \veps_l
            -(\bgamma_{j(i-1)}'\bgamma_{j(i-1)})^{-1}\sum_{k\in \cB_{j(i-1)}^0} k^2\sum_{l\in \cB_{j(i-1)}^0} \veps_l.
$$
Then $E [\tau_i]=0$ and ${\bf \Var}[\tau_i]\le 2\Delta^2$, where
\begin{equation}\label{kkt-2-1-6}
\Delta^2=\max_{i\in \cJ^0} \left\{
\left[(\bgamma_{j(i)}'\bgamma_{j(i)})^{-1}\sum_{k\in \cB_{j(i)}^0} k^2\right]^2 \sum_{l\in \cB_{j(i)}^0} l^2\right\} .
\end{equation}
Consider independent copies $\tau_i^*\sim N(0, 2\Delta^2)$. From \eqref{kkt-2-1-3} and the Slepian inequality, we have
\begin{equation}\label{kkt-2-1-7}
P(I_1^c)=P(\max_{i\in  \cJ_0}|\tau_i|>\frac{a_n}{2})\le P(\max_{i\in  \cJ_0}|\tau_i^*|>\frac{a_n}{2})\le \exp\{-\frac{a_n^2}{8\Delta^2}\}.
\end{equation}
From \eqref{kkt-2-1-6}, we know
$$
\Delta^2=\max_{i\in  \cJ_0}\left[\frac{\left(\sum_{k\in \cB_{j(i)}^0} k^2\right)^{3/2}}
{\left(\sum_{k\in \cB_{j(i)}^0} k\right)^2+\left(\sum_{k\in \cB_{j(i)}^0} k^2\right)^2}\right]^2
\le \max_{i\in  \cJ_0}\left(\sum_{k\in \cB_{j(i)}^0} k^2\right)^{-1}\le 3 / (b_{\min}^0)^3.
$$
The last ``$\le$'' is because $\sum_{k\in \cB_{j(i)}^0} k^2 \le |\cB_{j(i)}^0|^3/3$.
Thus $P(I_1^c)=0$ from (B2). Combining with \eqref{kkt-2-1-5}, we know that \eqref{kkt-2-1} holds with probability
to $1$ when $n\to \infty$.
In order to verify \eqref{kkt-2-2}, we consider the sub-differential on $\bmu$ vector,
\begin{equation}\label{kkt-2-1-8}
y_k-\hmu_k=\veps_k-(\hmu_k-\mu_k^0)=\lm_n(\widehat h_k)~ {\rm for~} 3\le k\le n,
\end{equation}
where $\widehat h_k=-2\sgn(\hmu_{k-1}+\hmu_{k+1}-2\hmu_k)+\sgn(\hmu_{k-2}+\hmu_{k}-2\hmu_{k-1})+\sgn(\hmu_{k}+\hmu_{k+2}-2\hmu_{k+1})$
for $3\le k\le n-2$, $\widehat h_n=\sgn(\hmu_{n-2}+\hmu_{n}-2\hmu_{n-1})$ and
 $\widehat h_{n-1}=-2\sgn(\hmu_{n-2}+\hmu_{n}-2\hmu_{n-1})+\sgn(\hmu_{n-3}+\hmu_{n-1}-2\hmu_{n-2}).$
If we apply  \eqref{kkt-2-1-8}
to $k=i-1, i$ and $i+1$ separately and then we get
$$
\veps_{i+1}+\veps_{i-1}-2\veps_i-[(\hmu_{i+1}+\hmu_{i-1}-2\hmu_i)-(\mu^0_{i+1}+\mu^0_{i-1}-2\mu^0_i)]
=\lm_n[\widehat h_{i+1}+\widehat h_{i-1}-2\widehat h_{i}].
$$
In fact,  $\widehat b_{j(i)}=\widehat b_{j(i-1)}$, or equivalently $i\notin \cJ^0$, means
$\hmu_{i+1}+\hmu_{i-1}-2\hmu_i=\mu^0_{i+1}+\mu^0_{i-1}-2\mu^0_i=0$.
Then we  \eqref{kkt-2-2} holds if
$$
|\veps_{i+1}+\veps_{i-1}-2\veps_i|\le 14\lm_n \quad {\rm for~} i\notin \cJ^0.
$$
Denote $d_i^{\bveps}=\veps_{i+1}/\sqrt{6}+\veps_{i-1}/\sqrt{6}-2\veps_i/\sqrt{6}$.
From (A1), $d_i^{\bveps}$ has  sub-Gaussian distribution with mean $0$ and variance $\sigma^2$.
Then
\begin{equation}\label{kkt-2-1-9}
E[\max_{i\notin\cJ^0} |d_i^{\bveps}| ]\le \frac{\sigma(2+\log(n-|\cJ^0|))}{2}.
\end{equation}
If $14\lm_n-(\sigma/2)(2+\log(n-|\cJ^0|))>0$, then from \eqref{kkt-2-1-9}, we have
\begin{equation}\label{kkt-2-1-10}
\begin{array}{ll}
P(\max_{i\notin\cJ^0}|d_i^{\bveps}|>14\lm_n)
&\le \exp\{-[14\lm_n-(\sigma/2)(2+\log(n-|\cJ^0|))]^2/2\\
&= \exp\{-(14\lm_n)^2[1-(\sigma/28)((2+\log(n-|\cJ^0|))/\lm_n)]^2/2\\
&\le \exp\{-(14\lm_n)^2 (1-\delta))^2\},
\end{array}
\end{equation}
where the first ``$\le$'' is from  the Borell-TIS inequality and
the second ``$\le$'' is from  (B3-b). 
Denote $I_3\equiv \{ \max_{i\notin\cJ^0}|d_i^{\bveps}|< 14\lm_n\}$.
Then $P(I_3^c)\to 0$ if (B3-a) holds. 
Thus
$\lim_{n\to \infty}P( \cS^{n})\ge 1-(P(I_1^c)+P(I_2^c)+P(I_3^c))=1.$ $\Box$

\newpage

 \noindent{\bf A counter-example for weak irrepresentable condition in Section \ref{sec-com}}

\noindent

Specifically, the design matrix of model \eqref{lasso} is
$$
\bZ=
\begin{pmatrix}
    1&      &       &      &    &    &      \\
    1&      1&      &      &    &    &      \\
    1&      2&      1&     &    &    &      \\
    1&      3&      2&     1&    &   &       \\
\vdots&\vdots&\vdots&\vdots&\ddots&     1&      \\
1&      n-1&    n-2&    n-3&\cdots&     2&      1
\end{pmatrix}.
$$
Suppose $t_1^0<t_2^0<\cdots<t_J^0$ are all
true kink points.
Denote $d^{n_k}=\sum_{i=1}^{n_k} i$ and
$d^{n_k}_{l,m}=\sum_{i=1}^{n_k}i(i+t_{j_m}-t_{j_l})$ with $n_k=n-(t_k-1)$ for $1\le k, l, m\le J$.
We can write $\bZ_{\bone}'\bZ_{\bone}$ explicitly. We as follows,
$$
\bZ_{\bone}'\bZ_{\bone}=
\begin{pmatrix}
   n                      & d^{n_1}       &d^{n_2}
 &d^{n_3}             &\cdots &d^{n_{J-1}}   & d^{n_{J}}      \\
d^{n_{1}} &d^{n_{1}}_{1,1} &d^{n_2}_{1,2}&d^{n_{3}}_{1,3}
 & \cdots & d^{n_{J-1}}_{1,J-1}    &  d^{n_{J}}_{1,J}\\
 d^{n_{2}} &d^{n_{2}}_{1,2} &d^{n_{2}}_{2,2}
& d^{n_{3}}_{2,3} &\cdots    &d^{n_{J-1}}_{2,J-1}
&d^{n_{J}}_{2,J}    \\
 d^{n_{3}} &d^{n_{3}}_{1,3} &  d^{n_{3}}_{2,3}
&d^{n_{3}}_{3,3}&\cdots    & d^{n_{J-1}}_{3,J-1}   &d^{n_{J}}_{3,J}     \\
  \vdots  &  \vdots & \vdots      &\vdots&\ddots&\vdots& \vdots    \\
d^{n_{J-1}} &d^{n_{J-1}}_{1,J-1} &d^{n_{J-1}}_{2,J-1}
 &d^{n_{J-1}}_{3,J-1}&\cdots    &d^{n_{J-1}}_{J-1,J-1} &d^{n_{J}}_{J-1,J}(i)    \\
 d^{n_{J}} &  d^{n_{J}}_{1,J}& d^{n_{J}}_{2,J}
&d^{n_{J}}_{3,J}&\cdots    & d^{n_{J}}_{J-1,J}   &d^{n_{J}}_{J,J}
\end{pmatrix}.
$$
Here $t_1=2$ since there is no penalty on $\beta_1$ and $\beta_2$, both $\bz_1$
and $\bz_2$ are included in $\bZ_{\bone}$.
Suppose there is only one counter example $t_1$.
$$
\bZ_{\bone}'\bZ_{\bone}=
\begin{pmatrix}
   n                      & \sum_{i=1}^{n-1} i       &\sum_{i=1}^{n-t_1+1} i     \\
 \sum_{i=1}^{n-1} i & \sum_{i=1}^{n-1}i^2 & \sum_{i=1}^{n-t_1+1}i(i+t_1-2)\\
\sum_{i=1}^{n-t_1+1} i   &\sum_{i=1}^{n-t_1+1} i  (i+t_1-2) &\sum_{i=1}^{n-t_1+1} i^2
\end{pmatrix}.
$$
Without loss of generalicity, we let $n=10$, $t_1=5$. Then the seven 3-d vectors in
$\bZ'_{\btwo}\bZ_{\bone}(\bZ'_{\bone}\bZ_{\bone})^{-1}$ are
$\ba_1=(-0.3255, 0.7383, 0.2872)'$, $\ba_2=(-0.2383,  0.3574, 0.6809)$, $\ba_3=(0.1277, -0.1915, 1.0638)'$
$\ba_4=(0.1702, -0.2553, 0.9422)'$, $\ba_5=(0.1532, -0.2298, 0.7052)'$, $\ba_6=(0.1021, -0.1532, 0.4225)'$
and $\ba_7=(0.0426, -0.0638, 0.1641)'$.
If $\bs_{\bone}=(1,1,1)'$, then $|\ba_j'\bs_{\bone}|=1$ for $j=3$.
If $\bs_{\bone}=(1,-1,1)'$, then $|\ba_j'\bs_{\bone}|>1$ for $j=3,4,5$.
If $\bs_{\bone}=(1,1,-1)'$, then $|\ba_j'\bs_{\bone}|>1$ for $j=3,4$.
If $\bs_{\bone}=(-1,1,1)'$, then $|\ba_j'\bs_{\bone}|>1$ for $j=1,2$.$\Box$

\newpage

\begin{table} [h] \label {table 1}
 \begin{center}
   \caption {Simulation results using SIC and MC for Example 1 and 2 in Section \ref{sec-sim}.
   }
  {\small  \begin{tabular} {c c c  c c c   c  c c c  }\label{sim results} \\ \hline \hline
                       &  &  &  \multicolumn{3}{c}  {$n=500 $}
                           & & \multicolumn{3}{c}  {$n=1000$}
              \\
                   \cline{4-6} \cline{8-10}
                      &  &    & Low& Medium& High&
                            &  Low& Medium& High
                \\ \hline
                              \multirow{17}{*}{\rotatebox{90}{\mbox{Example 1 ($k=3$)}}}\\
& &  RE     & 0.000 & 0.002   &0.039   &       &0.000   &0.002  & 0.047   \\
& &           &  (0.000) &(0.003) &(0.089) && (0.000) &(0.003)& (0.141) \\
 &  & $|\cJ|$   &20.83 & 31.14  & 41.19  & & 44.12  & 14.42  &10.59    \\
&   &         & (76.365) &(105.807) & (125.186)& &(166.980)& (7.643) & (6.612) \\
 & \raisebox{1.3ex}[0pt]{SIC}
  & eAB   & 0.003& 0.009  & 0.030        &    &    0.002   &0.008 & 0.029  \\
&  &  & (0.002)&(0.006) & (0.022) &&(0.001)& (0.005) &(0.019)\\
&  & eBA   & 0.286 & 0.261  & 0.249       &    &     0.275  &0.255  &0.239 \\
 & &   &(0.013)& (0.049)& (0.059)&& (0.032) &(0.056)& (0.067)\\
\\
& &   RE     & 0.000 & 0.003    &    0.058 & & 0.000        &0.003  &  0.069  \\
& &&(0.000)&(0.004)& (0.134) &&(0.000)&(0.005)&(0.177)\\
&   & $|\cJ|$   &3.54 & 3.66   & 3.68        && 8.01          & 7.55 &5.41    \\
&  & & (1.009)& (1.094)&(1.034)&&(1.811)& (1.855)&(1.450)\\
&  \raisebox{1.3ex}[0pt]{MC}
  & eAB   & 0.003  &  0.009 & 0.034     &   &      0.002    &0.009  & 0.032 \\
&  &&(0.002) &(0.006)& (0.023) && (0.001)&(0.005)&(0.020) \\
&  & eBA   &0.275 & 0.243 & 0.231   &        &     0.256      & 0.213 & 0.205 \\
&  && (0.028)&(0.055)&(0.061)&&(0.045)& (0.064)& (0.076)\\
  \\
   \multirow{17}{*}{\rotatebox{90}{\mbox{Example 2 ($k=5$)}}}\\
&  &   RE     & 0.015 &  0.009   &0.037   &    & 0.018  &0.015  & 0.034   \\
 & &          &  (0.000)& (0.003) &(0.047)&&(0.001) &(0.003)& (0.044)\\
&     & $|\cJ|$   & 5.89 & 7.07    & 11.25    &  &5.38  & 6.28   & 13.48  \\
&      &&  (1.377)&(1.565)& (4.368) & &(1.099)& (1.341) &  (5.668)\\
&  \raisebox{1.3ex}[0pt]{SIC}
  & eAB       & 0.003 &  0.010         &0.033 & &   0.003     &0.009  & 0.031\\
&     &  &(0.001) & (0.005) &(0.018)&& (0.001)&(0.005)& (0.021) \\
&  & eBA   & 0.186  &    0.038            & 0.165 &&    0.005   & 0.027  & 0.178 \\
&       &&(0.042) &(0.025) &(0.036) &&(0.003)& (0.015) &  (0.023)\\
\\
& &   RE     &0.016 &  0.011 &    0.047  &       & 0.019    &0.015&  0.047\\
&      && (0.002)  & (0.005)&(0.056) &&(0.001)&   (0.004)& (0.051)  \\
 &    & $|\cJ|$   & 5.38  & 6.03   & 6.48   &  &5.13      & 5.87  & 5.53  \\
&          &&(1.022) &(1.344) &  (1.629) & & (0.884) & (1.088)& (1.322) \\
&  \raisebox{1.3ex}[0pt]{MC}
  & eAB   & 0.013&   0.012& 0.049         &    & 0.003  &0.011 & 0.055 \\
&       && (0.043) & (0.019)&(0.043) &&(0.001)&  (0.019)&(0.053)\\
&  & eBA   & 0.186 & 0.033  & 0.133   &           & 0.005 & 0.026 & 0.132  \\
&       && (0.042)&(0.023)&(0.050) &&(0.003)& (0.015) &(0.055)\\
         \hline\hline
\multicolumn{10}{l} {\footnotesize{NOTE 1: RE is the   relative error  defined in \eqref{eq:lre}.}}\\
 \multicolumn{8}{l} {\footnotesize{NOTE 2:  eAB and eBA are  $\cE(A,B)$ and $\cE(B,A)$ in \eqref{eq:hd} divided by $n$.}}\\
\multicolumn{10}{l} {\footnotesize{NOTE 2:  $|\cJ|$ is defined estimated kink points number.}}\\
\multicolumn{10}{l} {\footnotesize{NOTE 3:  Values in the parenthesis are for corresponding standard deviations.}}
   \end{tabular}}
  \end{center}
  \end{table}

\begin{table} [h] \label {table 2}
 \begin{center}
   \caption {LASSO output using MC for Example 1 and 2 with $n=1000$ in Section \ref{sec-sim}.
   }
  {\small  \begin{tabular} { c  c c c  c  c c c  }\label{sim results} \\ \hline\hline
                          &  \multicolumn{3}{c}  {Example 1 ($k=3$)}
                          &  & \multicolumn{3}{c}  {Example 2 ($k=5$)}
              \\
                         \cline{2-4} \cline{6-8}
                            & Low& Medium& High
                          &  &  Low& Medium& High
                             \\  \hline
   RE        &0.024  &0.036  & 1.085 &   &1e-4          &1e-4& 0.001  \\
            & (0.003)&(0.011)&(0.289)&   &(1e-5)           & (1e-5) & (1e-6)\\
   $|\cJ|$         & 42.16 & 41.38 &51.94 &     &  91.26 & 92.56  &96.04    \\
                  & (1.434)& (4.899)& (14.816)&&  (0.443)&(3.494)&(1.470) \\
     eAB        &0.000  & 0.000&  0.001&     &0.000     & 0.001 &  0.014 \\
                &(0.000)& (0.000) & (0.002)&  & (0.000)& (0.001)&(0.000)\\
      eBA          & 0.067& 0.099 &0.293&   & 0.032   & 0.032  &0.036  \\
                & (0.019) & (0.053)&(0.003)&&  (0.000)&(0.002)&(0.000) \\
        \hline\hline
\multicolumn{8}{l} {\footnotesize{NOTE 1: RE is the   relative error  defined in \eqref{eq:lre}.}}\\
 \multicolumn{8}{l} {\footnotesize{NOTE 2:  eAB and eBA are  $\cE(A,B)$ and $\cE(B,A)$ in \eqref{eq:hd} divided by $n$.}}\\
\multicolumn{8}{l} {\footnotesize{NOTE 3:  $|\cJ|$ is defined estimated kink points number.}}  \\
\multicolumn{8}{l} {\footnotesize{NOTE 4: Values in the parenthesis are for corresponding standard deviations.}}
   \end{tabular}}
  \end{center}
  \end{table}

\end{document}